\documentclass[traditabstract]{aa} 
\usepackage{epsfig}
\usepackage{graphicx}
\usepackage{xcolor}
\usepackage{txfonts}
\usepackage{amssymb}
\usepackage{natbib}   
\usepackage{hyperref}

\bibpunct{(}{)}{;}{a}{}{,}
\newcommand{\be}{\begin{equation}}
\newcommand{\ee}{\end{equation}}
\newcommand{\bea}{\begin{eqnarray}}
\newcommand{\eea}{\end{eqnarray}}
\renewcommand{\vec}[1]{\mbox{\boldmath ${#1}$}}

\makeatletter
\def\fvec#1{\underline{\sbox\tw@{$#1$}\dp\tw@\z@\box\tw@}}
\makeatother

\usepackage[normalem]{ulem}

\begin{document}

\title{Expected polarization properties of nonmagnetized CCOs}
\author{
Valery~F. Suleimanov\inst{1}
\and
Juri Poutanen\inst{2}
\and
Victor Doroshenko\inst{1}
\and 
Klaus Werner\inst{1}
}


\institute{
Institut f\"ur Astronomie und Astrophysik, Kepler Center for Astro and Particle Physics, Universit\"at T\"ubingen, Sand 1,  72076 T\"ubingen, Germany\\ \email{suleimanov@astro.uni-tuebingen.de} 
\and Department of Physics and Astronomy,  FI-20014 University of Turku, Finland 
}

\date{Received 7 February 2023 / Accepted  2 March 2023}

\authorrunning{Suleimanov et al.}
\titlerunning{Polarization properties of CCOs}

\abstract
{Central compact objects (CCOs) are neutron stars found close to the center of some supernova remnants. A certain number of them are presumably  covered by carbon envelopes. Their unpulsed thermal X-ray emission can originate either from the entire surface covered by a carbon atmosphere or alternatively from a nonuniformly emitting hydrogen atmosphere. However, the latter scenario appears unlikely given the available upper limits on the amplitude of pulsations. 
Here we explore a possibility to further discriminate between the two scenarios using X-ray polarimetric observations. We compute the polarization degree (PD) for nonmagnetized pure-carbon and pure-hydrogen atmospheres with effective temperatures of between 1 and 6 MK and find that it can reach up to 25\% and 40\%  for hydrogen and carbon atmospheres, respectively, in the photon energy band 1--10 keV. 
However, given the available constraints on possible inhomogeneities of the temperature distribution deduced from models of the X-ray spectrum of the CCO in HESS J1731$-$347, the integrated PD appears to be very low for both carbon ($<0.25\%$) and hydrogen (a few percent) compositions in the energy band of 2--8 keV covered by the recently launched \textit{Imaging X-ray Polarimetry Explorer}. 
We therefore conclude that polarization from CCOs is not expected to be detectable by current facilities, but future detection would strongly support nonuniform hydrogen composition models.
}


\maketitle
%
\section{Introduction}

A neutron star (NS) is the product of gravitational collapse of the iron core of a massive star in the final stages of its evolution followed by a supernova explosion.
It is therefore not uncommon to find NSs at the centers of some supernova remnants. Some of these, the so-called central compact cbjects \citep[CCOs; see, e.g.,][]{Pavlov.etal:02, Pavlov.etal:04, deluca17}, appear to be radio-quiet, are presumably weakly magnetized, and are detectable only by their thermal X-ray emission.
Their X-ray spectra are well described by model atmospheres and they are attractive objects for investigations of NS interiors. Indeed, the distance to and the age of the host supernova remnant are usually known with reasonable accuracy, which is important for the investigation of supradense cold-matter properties in NS cores \citep[see, e.g.,][and references therein]{2016PhR...621..127L} and verification of NS cooling theory  \citep[see, e.g.][and references therein]{2004ARA&A..42..169Y,2015SSRv239P,2020MNRAS.496.5052P}. In particular, estimates of the observed NS radius are important for constraining equation of state of cold dense matter \citep[see, e.g., reviews by][]{2013ApJ...765L...5S,2016EPJA...52...20S,2018ASSL..457..185D}.

There are several CCOs that are probably covered with carbon envelopes because their observed spectra are well described with model spectra of pure-carbon atmospheres, and alternative compositions yield unreasonable NS parameters albeit formally consistent with the observed spectra. 
This was first suggested for Cas A  \citep{2009Natur.462...71H}, where a NS hydrogen-atmosphere model required an unreasonably small radius unless a nonuniform temperature distribution was considered over the surface. 
However, the latter option appears to be inconsistent with the absence of detectable X-ray pulsations \citep{2009ApJ...703..910P}. 
At present, we know of at least five more CCOs where carbon atmosphere composition is plausible; namely the CCOs in the supernova remnant HESS J1731$-$347 \citep{2013A&A...556A..41K}, as well as G15.9+0.2 \citep{2016A&A...592L..12K}, G330.2+1.0 \citep{2018A&A...618A..76D}, G347.3--0.5, and 
G350.1--0.3   \citep{2021MNRAS.506.5015H}.
It is not fully clear why their atmospheres could be pure carbon. 
We previously suggested \citep{2018A&A...618A..76D} that the simplest explanation is that the progenitors of these NSs were highly evolved stars with low hydrogen and helium abundances in their outer envelopes; for example, Wolf-Rayet stars. 
The supernova explosion then takes place in a wind-blown bubble devoid of hydrogen and helium, and so after explosion the hydrogen- and helium-free envelope can be retained.  

However, there is also a possibility that the surfaces of these CCOs are still covered by hydrogen and have a nonuniform temperature distribution (\citealt{2017A&A...600A..43S}, see also \citealt{2016ApJ...826..162E}). 
Indeed, even in this case, the observed emission can be unpulsed if the hotter regions are located close to the rotation axis, or if the line of sight is directed close to the rotation axis. 
Nevertheless, the probability of such an unfavorable geometry for all six CCOs where carbon atmosphere has been suggested is very low \citep{2021RAA....21..294W}. 
On the other hand, it cannot be completely excluded for at least some individual objects, a spectacular example being the CCO in Kes~79, which has the largest pulsed fraction among all known CCOs, namely $\approx$\,64\%. 
The high observed pulsed fraction implies a relatively small size for the emitting hot spots, but at the same time the pulse-phase-averaged spectrum of this CCO is well fitted by carbon atmosphere spectra with the emission area comparable to the entire NS surface \citep{2014ApJ...790...94B}.
Therefore, the problem of chemical composition of CCO atmospheres may clearly be more complex than it appears at first sight.

For some objects, such as the CCO in HESS J1731$-$347, finding an unambiguous solution to this problem may be especially relevant. 
Recently, \citet{2022NatAs.tmp..224D} found that even under the assumption of a pure-carbon atmosphere, the NS needs to be relatively compact and light. 
In particular, the mass of $\sim0.8M_\odot$ was estimated assuming a uniformly emitting carbon atmosphere, which opens up the intriguing possibility that the CCO might actually be a more exotic object, such as a strange (quark) star. 
As in the case of other CCOs, the central object could also be a NS with more standard parameters if a nonuniform temperature distribution over the surface is assumed and the pulsations are not detectable due to an unfavorable viewing geometry. 

Here, we try to clarify whether or not the geometry and chemical composition of the NS atmosphere can be constrained using an independent observational window opened with the launch of the \textit{Imaging X-ray Polarimetry Explorer} (\textit{IXPE}) in 2021 \citep{Weisskopf2022}. 
Indeed, while a uniformly emitting NS surface is not expected to produce any observable polarization, emission coming from compact hot spots covered by an atmosphere may exhibit some polarization. 
We study the local polarization properties of carbon and hydrogen NS atmospheres and compute the expected polarization of nonuniform NS models with various chemical compositions in conjunction with parameter settings that reproduce the observed spectrum of the CCO in HESS J1731$-$347 \citep{2022NatAs.tmp..224D}. 
We find that the resulting PD is, in any case, relatively low when integrated over the NS surface and unlikely to be detected with current instrumentation. 
On the other hand, a detection of the polarization from this CCO could still be feasible in the future and would strongly favor the nonuniform hydrogen model.
 
\section{Method}
\label{sec:method}

\begin{figure}
\centering
\includegraphics[width=0.85\columnwidth]{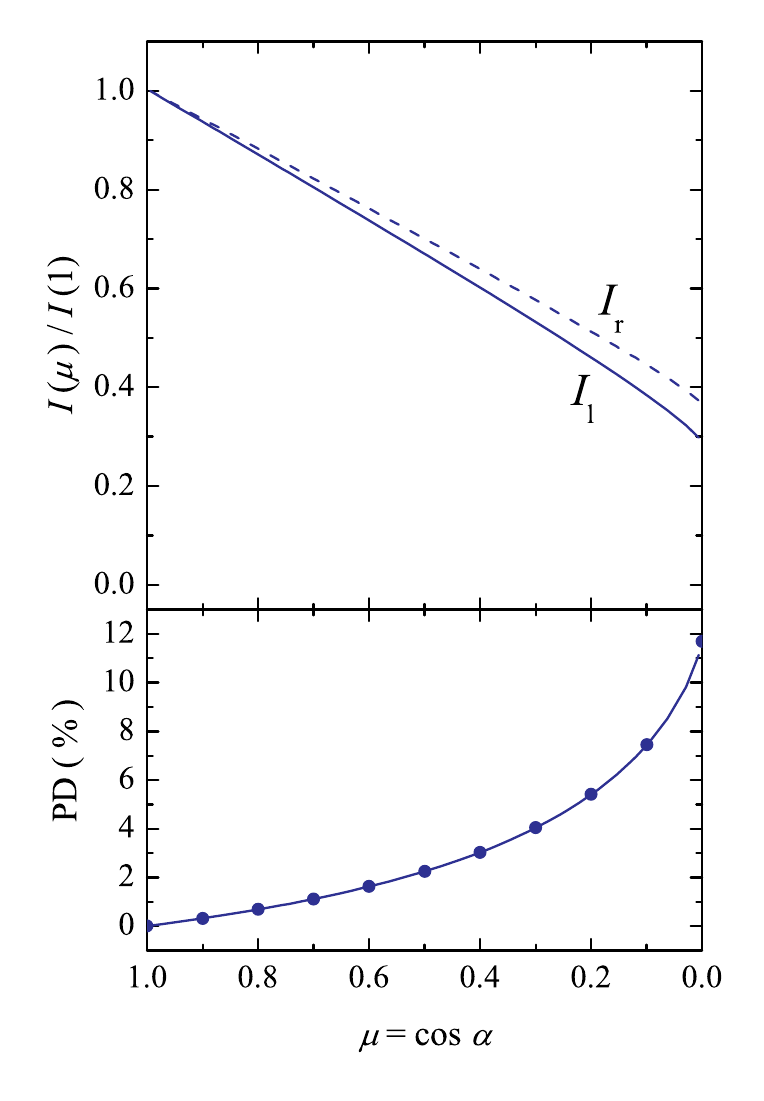}
\caption{Angular dependences of the specific intensities in two polarization modes (\textit{top panel}) and PD (\textit{bottom panel}) for pure-electron-scattering semi-infinite atmosphere. 
Lines correspond to our calculations. 
Solid and dashed lines in the top panel are for $I_l$ and $I_r$, respectively. 
Circles at the bottom panel are the exact Chandrasekhar-Sobolev solution \citep{Chandra1960,Sobolev63}. 
}
\label{fig:epol}
\end{figure}

\subsection{Calculation of local polarization}

Considering that the magnetic fields of CCOs are believed to be low, we ignore possible existing magnetic fields in CCO atmospheres, consider electron scattering as the only source of polarization, and ignore circular polarization. 
The radiation field depends on the photon energy and cosine of the zenith angle $\mu=\cos\alpha$ but is independent of the azimuth, resulting in a Stokes $U$ parameter being equal to zero. 
Thus, the radiation field can be fully described by two intensities $I_l$ and $I_r$, which represent two independent streams of oppositely polarized light: one in the meridional plane formed by the normal to the local surface and the photon momentum, and one perpendicular to that plane, respectively \citep[see][]{Chandra1960}.
The model of the atmosphere depends very little on polarization; therefore, to compute polarization properties we take the precomputed atmosphere models \citep{2014ApJS..210...13S}. 
We then solve the radiative transfer equation written here for the vector $\tilde{I}=
   \left( 
   \begin{array}{c} I_{l} \\ I_{r} \end{array}
   \right)$
\citep{Chandra1960,Sobolev63}: 
\bea \label{eq:RTE}
   \mu\frac{d\tilde{I}_{E}(\mu)}{d\tau}   
   &=& \frac{1}{\lambda_{E}} \,
      \tilde{I}_{E}(\mu) - 
   \left( \frac{1}{\lambda_{E}}-1 \right) \frac{B_E}{2} 
   \left( 
   \begin{array}{c} 1 \\ 1 \end{array}
   \right)  \\  
   & -& \frac{3}{8} 
 \int_{-1}^{+1}  
    \left( 
   \begin{array}{cc} 
   2(1-\mu^2)(1-\mu'^2)+\mu^2\mu'^2 & \mu^2 \\
   \mu'^2 &  1 \\
   \end{array}
   \right)   
 \tilde{I}_{E}(\mu') \,d\mu'\ , \nonumber 
 \eea   
where $\lambda_{E}=\kappa_{\rm sc}/(\kappa_{\rm sc}+\kappa_E)$ is the energy-dependent albedo for single scattering, that is, the ratio of the scattering to the total opacity. 
The  true opacity $\kappa_E$ arises mainly due to bremsstrahlung and photoionization. 
The optical depth here is defined for the electron scattering only, because it does not depend on energy, that is, 
\be  \label{eq:tau}
         d\tau = - \kappa_{\rm sc} dm,
\ee
with $\tau=0$  at the atmosphere surface   and $m$ being the column density.
Sources of thermal radiation are described by the Planck function $B_E$. 
For the precomputed atmosphere model, we already have the dependencies of $\lambda_{E}$ and $B_E$ on $\tau$. 

This system of equations is solved by the Feautrier method \citep{1964SAOSR.167...80F} with the standard boundary conditions for stellar atmospheres, namely the absence of  irradiation at the upper boundary,
\be
\tilde{I}_{E}(\tau=0,-\mu) =0,\quad \mu>0,  
\ee
and thermodynamic equilibrium at the bottom of the atmosphere  for each polarization mode:
\be
\tilde{I}_{E}(\tau=\tau_{\rm max},\mu) =\frac{B_E}{2}    
\left( 
\begin{array}{c} 1 \\ 1 \end{array}
\right)   .
\ee

To test our code, we first solved the radiative transfer equation (\ref{eq:RTE}) for an almost pure electron-scattering atmosphere, assuming 
$\kappa_E/\kappa_{\rm sc}=10^{-7}$, that is, $\lambda_{E}\approx 1- 10^{-7}$. 
The results shown in Fig.\,\ref{fig:epol} are in good agreement with the results presented in \citet{Chandra1960}.
The PD is defined here as
\be  \label{eq:pd}
\mbox{PD} = \frac{I_r - I_l}{I_r+I_l},
\ee 
that is, it is positive when the electric vector is predominantly  perpendicular to the meridional plane.  

\begin{figure}
\centering
\includegraphics[width=0.85\columnwidth]{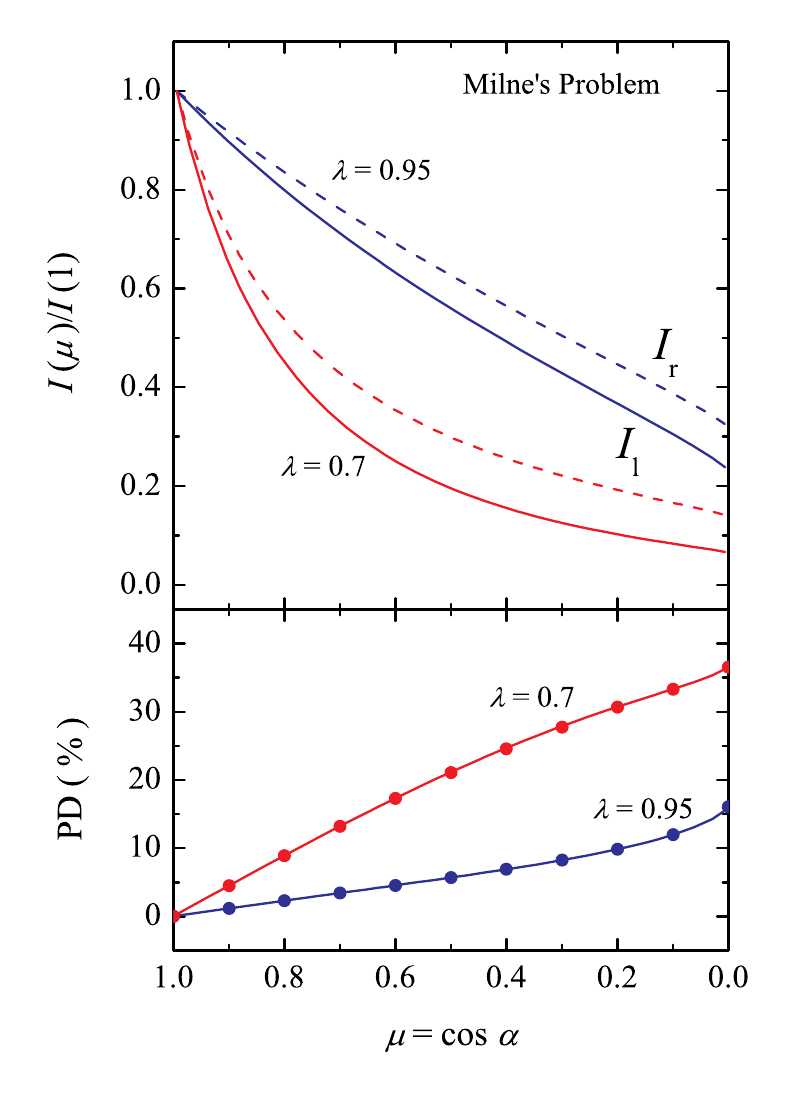}
\caption{Dependence of the specific intensities in two polarization modes ({\it top panel}) and PD ({\it bottom panel}) on the zenith angle for the atmosphere without sources of photons (Milne's problem, $B=0$). 
Blue and red curves correspond to our calculations for $\lambda_{E}=0.95$ and 0.7, respectively. 
Solid and dashed lines it the top panel are for $I_l$ and $I_r$, respectively. 
Circles at the bottom panel are the solution presented in \citet{1979Ap.....15..162L}. 
}
\label{fig:LS79a}
\end{figure}

\begin{figure}
\centering
\includegraphics[width=0.85\columnwidth]{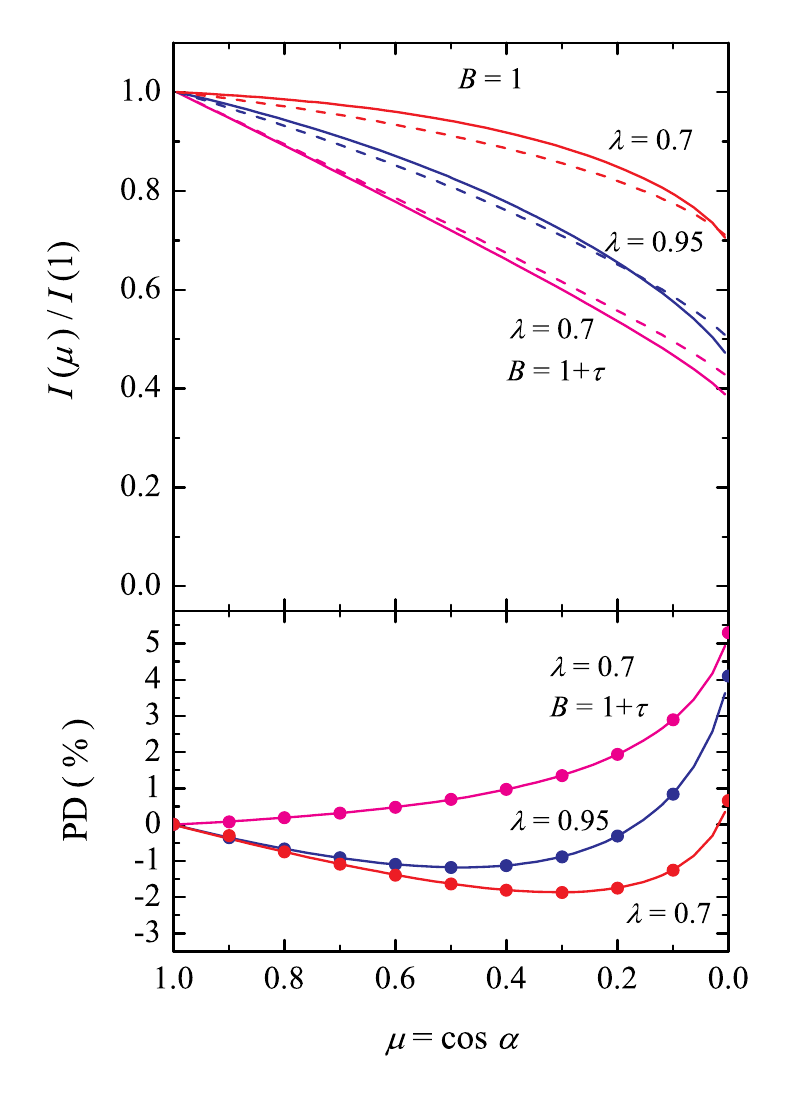}
\caption{Dependence of the specific intensities in two polarization modes ({\it top panel}) and PD ({\it bottom panel})  on the zenith angle for the  atmosphere with constant sources of photons over depth ($B=1$). 
Blue and red curves correspond to our calculations for $\lambda_{E}=0.95$ and 0.7, respectively.  
Solid and dashed lines in the top panel are for $I_l$ and $I_r$, respectively. 
The case where  $B=1+\tau$ and $\lambda_{E}=0.7$ is also shown with magenta curves. 
Circles at the bottom panel are the solution presented in \citet{1979Ap.....15..162L}. 
}
\label{fig:LS79b}
\end{figure}

We also checked our code against the results obtained by \citet{1979Ap.....15..162L} for atmospheres with a fixed value of $\lambda_{E}$ and various distributions of the
photon sources over the atmosphere: (1) no photon sources in the atmosphere (Milne's problem; all the photons come
through the bottom boundary, $B = 0$); (2) a constant source of photons over the depth ($B = 1$);
and (3) a linear dependence of the photon sources over optical depth  ($B = 1+\tau_{\rm total}$, where
$\tau_{\rm total}=\tau/\lambda_{E}$).
Our results are a close match to those of \citet{1979Ap.....15..162L}; see Figs.\,\ref{fig:LS79a} and \ref{fig:LS79b}.
In some models with $B=1+\tau$ and $B=1,$ the PD becomes negative (i.e., electric vector predominantly in the meridional plane) at some angles and $\lambda_{E}$ values as discovered by D.I. Nagirner in 1962 \citep{Dolginov79}.  
The solutions  for Milne's problem at relatively low $\lambda_{E} \approx 0.6-0.8$ exhibit high PD. 
This effect is related to the strong beaming of radiation along the normal to the atmosphere (see Fig.\,\ref{fig:LS79a}). 
Scattering of this radiation in the surface layers produces high polarization perpendicular to their initial direction, that is, parallel to the surface. 

\subsection{Polarization properties integrated over the surface}

\begin{figure}
\centering
\includegraphics[width=0.85\columnwidth]{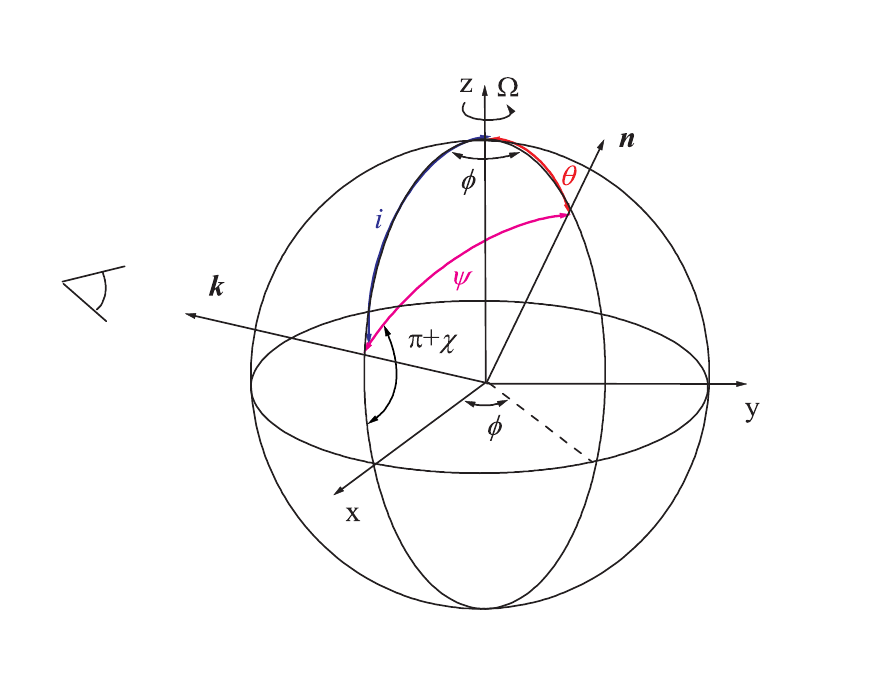}
\caption{Geometry of the problem. 
}
\label{fig:geom}
\end{figure}

Here, we consider a slowly rotating spherical NS with mass $M$ and radius $R$ and use the formalism developed in \citet{Viironen04}. 
We use a coordinate system defined by the NS rotation axis and the plane passing through the rotation axis and the line of sight as illustrated in Fig.\,\ref{fig:geom}. Here, $i$ is the inclination angle between rotation axis and line of sight, $\theta$ is the co-latitude of the considered point, $\phi$ is the rotational phase, and 
$\psi$ is the angle between the local normal $\vec{n}$ and the direction of the escaping photon in a flat space-time $\vec{k}$:
\be
       \cos\psi = \cos i \cos\theta + \sin i \sin\theta \cos\phi.
\ee

To obtain the Stokes parameters as observed by a distant observer, we have to integrate the local Stokes parameters over the observed part of the NS surface \citep{Viironen04}: 
\begin{eqnarray} \label{eq:FluxTotal}
\left( \begin{array}{c}
F_{E} \\ Q_{E} \\ U_{E} 
\end{array} \right)
& =& \frac{R^2}{D^2}  \int\! \! d\cos\theta \! \!  \int\! \!   d\varphi  \cos \alpha\ {\cal D}
\left(1+z\right)^{-3} \nonumber \\ 
&\times & I_{E '} (\alpha)
\left( \begin{array}{c}
 1 \\ \mbox{PD}_{E '}(\alpha) \cos 2(\chi_0+\chi) \\ \mbox{PD}_{E '}(\alpha)\sin  2(\chi_0+\chi)  
\end{array} \right) ,
\end{eqnarray} 
where $z$ is the gravitational redshift defined as
\be
  1+z = ( 1-u )^{-1/2} \quad \mbox{and} \quad u = \frac{2GM}{Rc^2},
\ee
$D$ is the distance to the NS, and $\alpha$ is the angle between the local normal and the direction of the escaping photons. 
This angle is smaller than the corresponding angle in flat space-time $\psi$ because of the light bending. 
Here, $\chi_0$ is the PA of the emitted radiation in the local basis, which is 
defined by the plane crossing the surface normal and the line of sight.
In our case of electron scattering, $\chi_0=90\degr$.
We use an accurate approximation for the relation between $\alpha$ and $\psi$ as proposed
by \citet{2020A&A...640A..24P}, where an analytical approximate formula for the lensing factor
\be
{\cal D} =\frac{1}{1-u}\frac{d \cos \alpha}{d \cos \psi} 
\ee
was also given. 

\begin{figure}
\centering
\includegraphics[width=1.0\columnwidth]{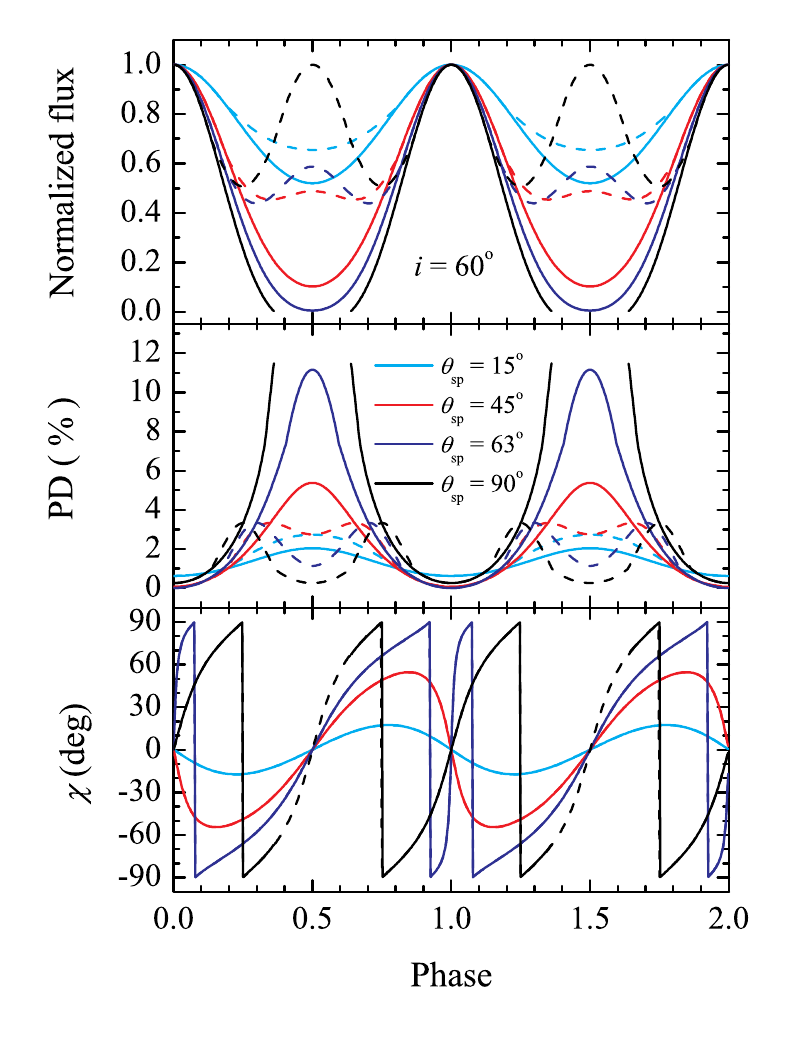}
\caption{Normalized monochromatic pulse profiles ({\it top panel}), PD ({\it middle panel}) and PA 
({\it bottom panel}) as a function of the rotation phase of a slowly rotating NS with
$M = 1.5\, M_\odot$ and $R = 12$\,km, and one (solid curves) or two antipodal (dashed curves) small bright spots
on the surface. The radiation from the spots has angular and polarization distributions similar to those of a pure electron-scattering atmosphere.
The computations were performed for inclination $i = 60\degr$ and four different spot co-latitudes  $\theta_{\rm sp} =$\,15\degr\,(light blue curves), 45\degr\,(red curves), 63\degr\,(blue curves), and 90\degr\,(black curves). 
This figure is similar to Fig.~11 from \citet{Viironen04} who considered a NS rotating at 400 Hz. 
}
\label{fig:sp_pol}
\end{figure}

It is also necessary to note that the integration in Eq. (\ref{eq:FluxTotal}) is performed over the  observable part of the surface  with $\cos \alpha > 0$. 
The values of $I_{E'}(\alpha, \theta, \varphi)$ and PD$_{E'}(\alpha, \theta, \varphi)$ depend on the local coordinates on the NS surface $\theta$ and $\varphi$--- because we consider a nonuniform temperature distribution--- and also on the angle $\alpha$. 
Now we measure the PA relative to the main polarization basis comprising the NS rotation axis and the line of sight. The PA measured in this frame is expressed using the rotating vector model \citep{1969ApL.....3..225R, 2020A&A...641A.166P}, as
\be
\tan \chi = \frac{\sin\theta \sin\phi}{\cos i \sin\theta \cos\phi -\sin i \cos\theta}. 
\ee

The final observed PD$_E$ and the PA$_E \equiv \chi_E$ are determined using
the total surface-integrated Stokes parameters
\be
\mbox{PD}_E = \frac{\sqrt{Q_E^2+U_E^2}}{F_E}, \qquad \tan 2\chi_E = \frac{U_E}{Q_E}.
\ee

Fortunately, the radiation transport in the gravitational field of a slowly rotating, spherically symmetric NS does not rotate the polarization plane of the emitting radiation. 
We note that this is not correct for rapidly rotating NSs or for accretion disks \citep[see, e.g.,][]{1980ApJ...235..224C, 2020A&A...641A.166P, Loktev20, 2022A&A...660A..25L}. 
Another favorable circumstance is the weak magnetic fields of the considered NSs. 
In the opposite case, we would have to consider the rotation of the polarization plane because of the vacuum birefringence in a strong magnetic field 
\citep[see, e.g.,][]{2003MNRAS.342..134H}.

We computed examples of pulse profiles and observed PD and PA from a slowly rotating NS with $M = 1.5\,M_\odot$ and $R = 12$\,km with one and two antipodal small bright spots. 
We assume that the local radiation of the bright spots has the same angular and polarization distributions as a pure electron-scattering atmosphere \citep{Chandra1960}. 
The computations were performed for an inclination $i = 60\degr$ and four co-latitudes of the bright spot $\theta_{\rm sp} =$\,15\degr, 45\degr, 63\degr, and 90\degr. 
The results presented in Fig.\,\ref{fig:sp_pol} are similar to those from \citet{Viironen04} who considered a rapidly rotating NS.  
The maximum polarization, up to 11.7\%, is expected from the one spot observed close to the NS edge ($\alpha \sim 90\degr$), but the observed flux is very low at those rotational phases. 
The expected polarization from two spots does not exceed a few percent and again is highest at the flux minima.

\section{Results}
\subsection{Local polarization for carbon atmospheres}

We also studied the local polarization properties of radiation emerging from a carbon model atmosphere using a test model with $T_{\rm eff} = 2$\,MK and $\log g = 14.3$ that was previously computed for a model atmosphere grid \citep{2014ApJS..210...13S}. 
We note that the model atmospheres were computed following a standard self-consistent approach and the extensively modified Kurucz's code \textsc{atlas} \citep{K70}. 
The number densities of various carbon ions were computed in all the atmosphere points using a local thermodynamic equilibrium (LTE) approximation. 
In the considered model atmospheres, carbon is not fully ionized and its hydrogen-like ion is responsible for the absorption edge at 0.45\,keV.

\begin{figure}
\centering
\includegraphics[width=0.85\columnwidth]{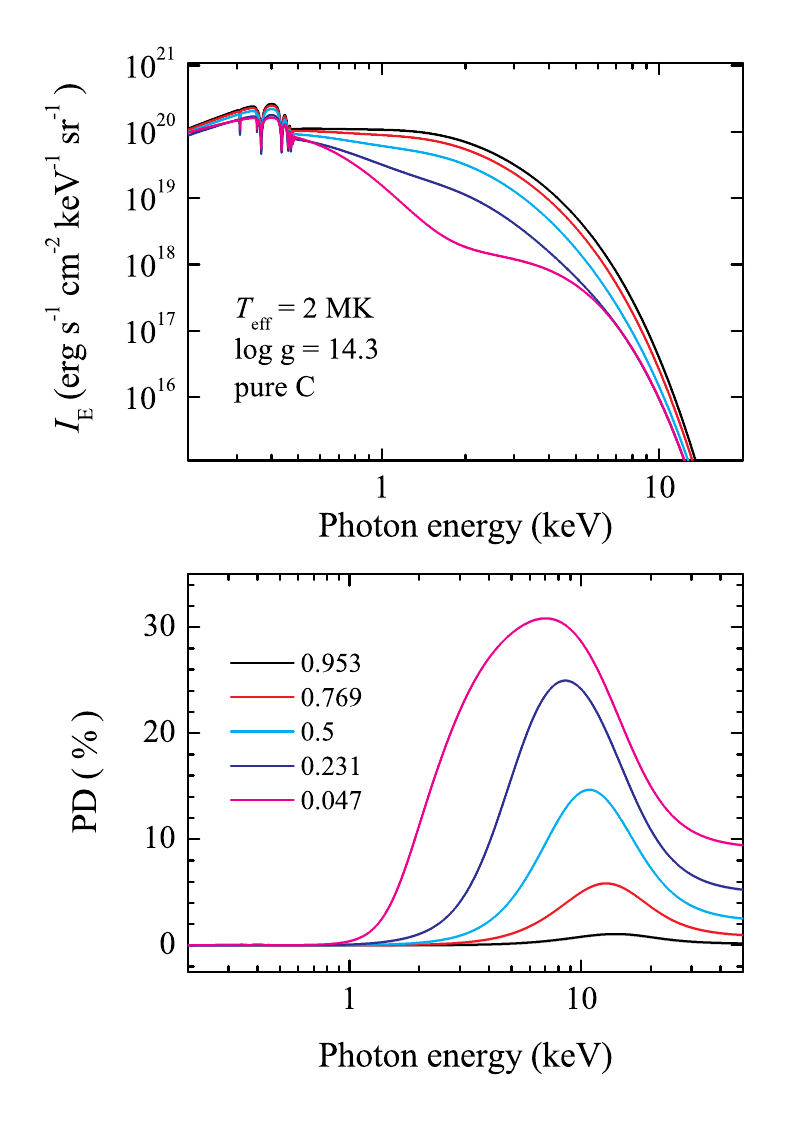}
\caption{Polarization properties of radiation emitted by pure-carbon atmospheres.
{\it Top panel:} Total specific intensity for $T_{\rm eff}= 2$\,MK and $\log g = 14.3$ at various angles to the normal: $\mu = 0.953$ (black), $0.769$ (red), $0.5$ (light blue), $0.231$ (blue), and $0.047$ (magenta).
{\it Bottom panel:} Corresponding energy dependence of the PD for the same five angles. 
}
\label{fig:catm2mk}
\end{figure}

The emergent spectra of total ($I_{r}+I_{l}$) specific intensity  for five angles between the photon transport direction and the normal $\alpha$, namely for $\mu = \cos \alpha =$ 0.953, 0.769, 0.5, 0.231, and 0.047, are shown in the top panel of Fig.\,\ref{fig:catm2mk}. 
The PD  of the emergent radiation for the same five angles along the spectral band are shown in the bottom panel of Fig.\,\ref{fig:catm2mk}.

\begin{figure}
\centering
\includegraphics[width=0.85\columnwidth]{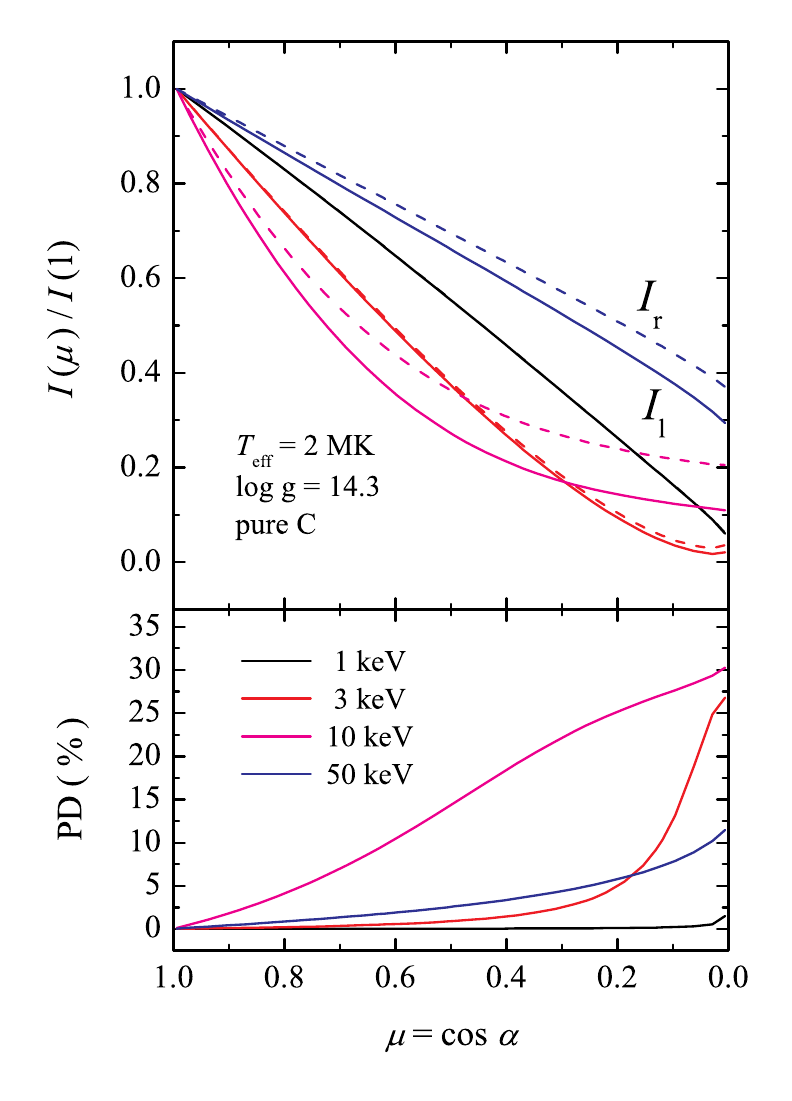}
\caption{Dependence of the specific intensities in two modes ({\it top panel}) and of the PD ({\it bottom panel}) on the angle to the normal for the same model atmosphere as in Fig.\,\ref{fig:catm2mk} for four photon energies, i.e., 1, 3, 10, and 50 keV, shown by black, red, magenta, and blue curves, respectively.  
Black solid and dashed curves coincide.
}
\label{fig:catm2mkmu}
\end{figure}

\begin{figure}
\centering
\includegraphics[width=0.85\columnwidth]{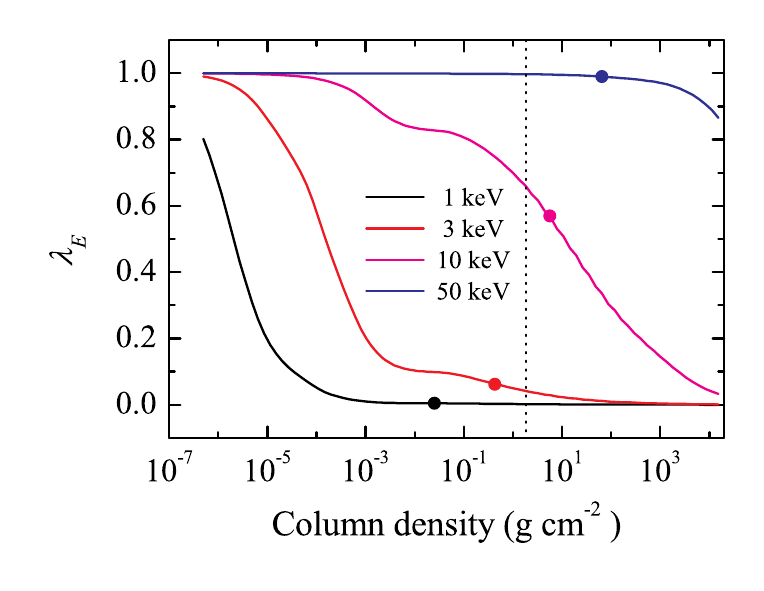}
\caption{Dependence of parameter $\lambda_{E}$, computed for the same energies and atmosphere model as in Fig.\,\ref{fig:catm2mkmu}, on the atmosphere depth.
Circles correspond to the approximate depth of the birth place of escaping photons ($\tau_{{\rm eff,}E}=1$). 
The vertical dotted line marks the approximate last-scattering depth ($\tau_{\rm sc}\approx 0.5$).
}
\label{fig:catm2mkalp}
\end{figure}

\begin{figure}
\centering
\includegraphics[width=0.85\columnwidth]{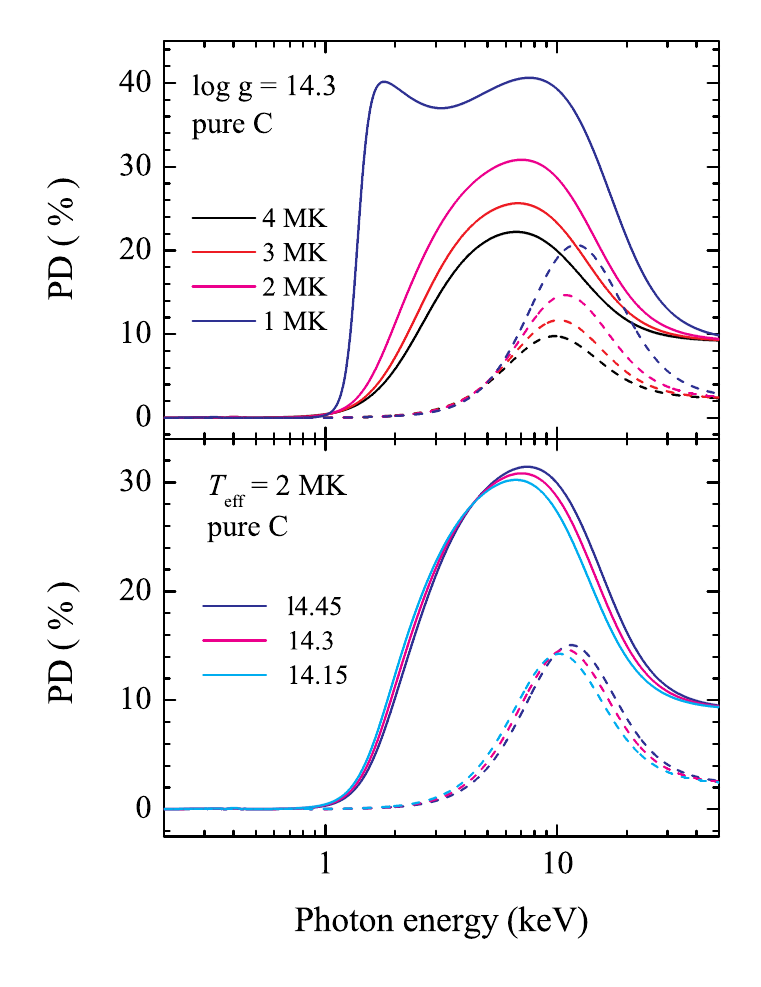}
\caption{Dependence of the PD on photon energy for two angles,
$\mu = 0.047$ (solid curves) and $0.5$ (dashed curves), for various effective temperatures
of the carbon atmospheres with $\log g = 14.3$ ({\it top panel}) and various surface gravities
with $T_{\rm eff} = 2$\,MK ({\it bottom panel}). 
}
\label{fig:catmvar}
\end{figure}

The PD is close to zero at photon energies below 1\,keV because the contribution of electron scattering to the total opacity is insignificant at these energies. 
At higher photon energies, the PDs increase up to their maxima at 10\,keV and further decrease to the values typical for a pure electron-scattering atmosphere.
These results can be understood using the polarization estimates presented above for Milne's problem. 
The emergent radiation is increasingly peaked along the normal at photon energies where the PD is highest; see the top panel of Fig.\,\ref{fig:catm2mk} \citep[see also Fig.\,3 in][]{2017A&A...600A..43S}. 
At the same photon energies, the contribution of electron scattering to the total opacity becomes dominant, $\lambda_{E} > 0.5$.   
In such conditions, the escaping photons scatter a few times after their birth. 
This means that the physical conditions close to Milne's problem provide a high PD, which exceeds that for a pure electron-scattering atmosphere. 
Electron scattering dominates at high photon energies, giving rise to $\lambda_{E} \approx 1,$ as in a pure electron-scattering-dominated atmosphere.

The interpretation proposed above is further confirmed by the plots shown in Figs.\,\ref{fig:catm2mkmu} and \ref{fig:catm2mkalp}. 
The top panel of Fig.\,\ref{fig:catm2mkmu} demonstrates that the emerging flux is much more peaked along the normal at the photon energies where the largest polarization was computed, namely at 3 and 10\,keV.
The angular distribution of the emerging radiation at 50\,keV almost coincides with that computed for a pure electron-scattering atmosphere (however, we note that for the temperatures observed for CCOs, there are almost no photons in this band.
The angular distribution of the radiation at 1\,keV is more relevant and is also peaked along the normal. 
However, the free-free and photoionization processes completely dominate over the electron scattering in this energy range, which makes the resulting polarization negligible. 

The bottom panel of the same figure shows the distributions of the PD over the zenith angle at the same four photon energies. 
As expected, the polarization at 1\,keV is close to zero and the polarization at 50\,keV coincides with the polarization dependence of a pure electron-scattering atmosphere.
The polarization dependence computed for 10\,keV is close to the solution obtained by \citet{1979Ap.....15..162L} for Milne's problem at $\lambda_{E} \approx 0.8$, whereas the polarization dependence computed for 3\,keV is low at almost all angles and exhibits a significant increase at large angles ($\mu < 0.2$).

Figure\,\ref{fig:catm2mkalp} helps us to understand the obtained behavior of the polarization at the presented photon energies.
The dependencies of $\lambda_{E}$ on the atmosphere depth are also shown. The positions of the birthplaces of 
the normally escaping photons, determined by the condition $\tau_{\rm eff,E}=1$, are shown with the circles. The effective 
optical depth is defined as
\be
  d\tau_{\rm eff,E} = \sqrt{\kappa_{\rm E}(\kappa_{\rm sc}+\kappa_{\rm E})}\,dm.
\ee
The position of the last scattering is indicated by the vertical dotted line. As expected, $\lambda_{E}$
is close to 1 at 50\,keV, and close to 0 at 1\,keV at almost all depths. Indeed, we see that the last scattering depth
is above the photon birthplace at photon energies of 10\,keV, and therefore the physical picture of the polarization forming is close
to Milne's problem. Moreover, the average value of $\lambda_{E}$ is close to 0.8 above the photon birthplace.
The picture is different for photons at 3\,keV. Most of the normally escaping photons are not scattering,
and the resulting polarization is low at large $\mu$. However, the photons escaping along the surface are scattering
a few times because $\lambda_{E}$ is large at the surface layers. This leads to the high PD at low $\mu$.
The same effect is weakly seen for the PD at 1\,keV.

The PD also depends on the parameters of the model atmosphere (see Fig.\,\ref{fig:catmvar}), and for example has the highest values for the lowest effective temperatures. However, at the same time, the flux level is the lowest at photon energies higher than 1\,keV  \citep[see, e.g., Fig.\,3 in][]{2014ApJS..210...13S}. 
The dependence of PD on the surface gravity is weak.    
  
\subsection{Local polarization for hydrogen atmospheres}

\begin{figure}
\centering
\includegraphics[width=0.85\columnwidth]{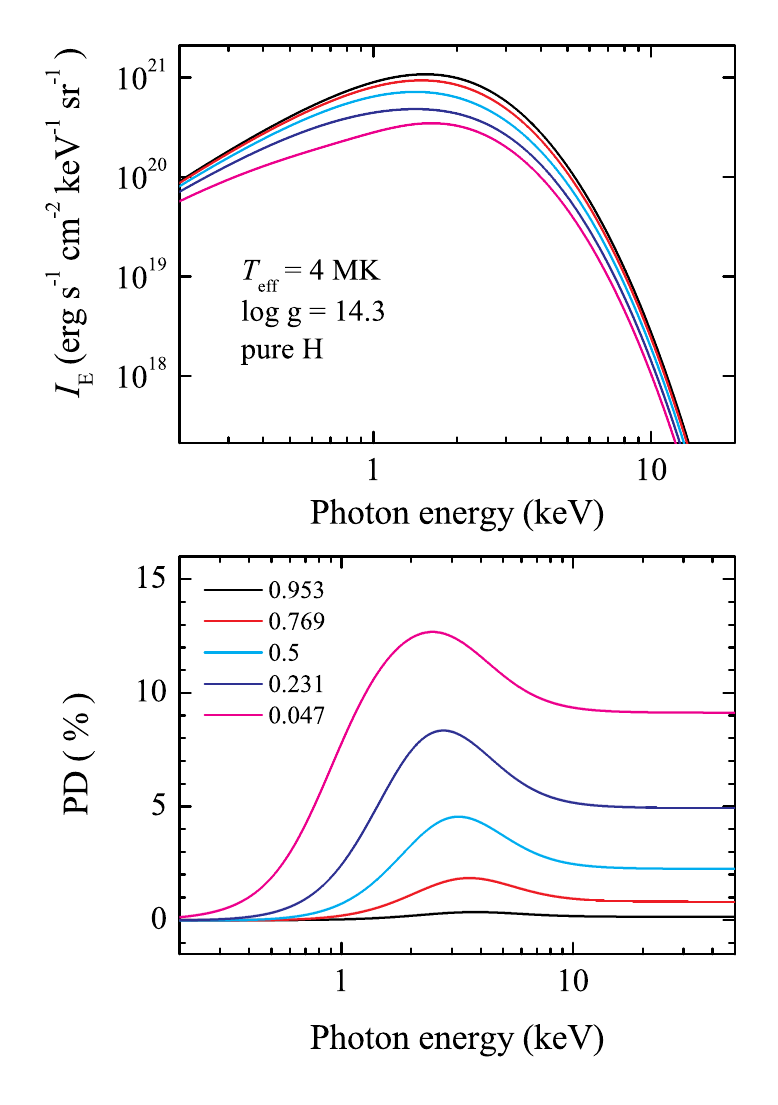}
\caption{Polarization properties of radiation emitted by pure-hydrogen atmospheres.
{\it Top panel:} Total specific intensity for the hydrogen atmosphere with $T_{\rm eff}= 4$\,MK and $\log g = 14.3$ at various angles to the normal.
The black, red, blue, light blue, and magenta curves correspond to $\mu = 0.953$, 
$0.769$, $0.5$, $0.231$, and $0.047$, respectively.
{\it Bottom panel:} Corresponding energy  dependence of the PD for the same five angles.
}
\label{fig:hatm4mk}
\end{figure}

\begin{figure}
\centering
\includegraphics[width=0.85\columnwidth]{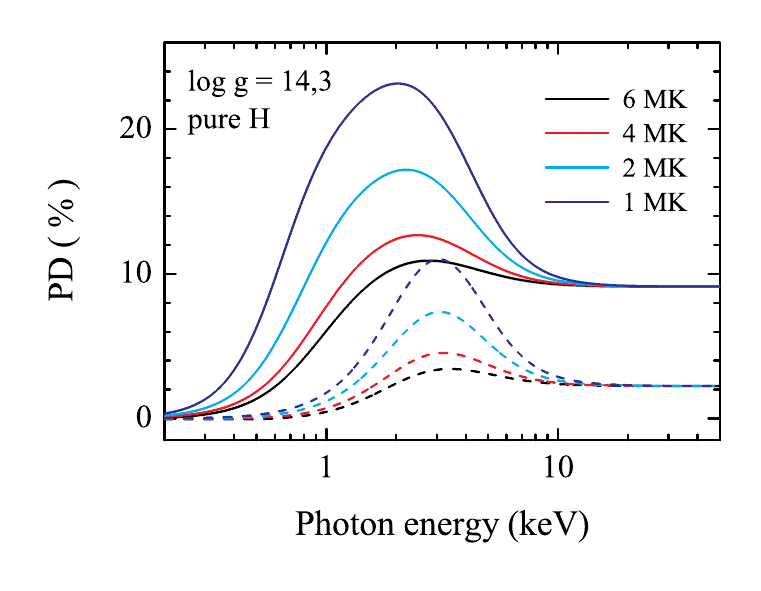}
\caption{Dependence of the PD on photon energy for two angles,
$\mu = 0.047$ (solid curves) and $\mu = 0.5$ (dashed curves) for various effective temperatures of the hydrogen atmospheres with $\log g = 14.3$. 
}
\label{fig:hatmvar}
\end{figure}

Polarization of the hydrogen model atmosphere spectra is qualitatively similar to 
that of the carbon model atmosphere spectra (see Figs.\,\ref{fig:hatm4mk} and 
\ref{fig:hatmvar}), although hydrogen is fully ionized in the considered
 atmospheres. The polarization is highest in the energy range of
0.5$-$5 keV, where a transition between the free-free opacity and electron scattering
occurs.  The value of $\lambda_{E}$ increases from a very low value to approximately 1, and the PD is close to the maximum possible values in those conditions, as discussed in the previous subsection.
At photon energies larger than 10\,keV, the polarization is close to the polarization properties of the pure electron-scattering atmosphere. However, the PDs of the hydrogen model atmosphere spectra are lower than in the spectra 
of carbon model atmospheres and do not even reach 25\,\% for the coolest model. The peak in the PD is also shifted to lower energies compared to the carbon atmospheres.    

\subsection{Polarization properties integrated over the surface for atmosphere models}

To estimate the observable polarization  from a distant NS  whose temperature is not uniform over the surface, we consider constraints on the geometry and temperatures of the two symmetric hot spots obtained based on analyses of X-ray spectra for  CCO in  HESS J1731$-$347 \citep{2022NatAs.tmp..224D} for hydrogen and carbon atmosphere models. The parameters of these two models are listed in Table\,\ref{tab1}.
We fixed the effective temperatures  equal to the previously  computed grid model values \citep{2014ApJS..210...13S,
2017A&A...600A..43S}. 
The parameter $f_1$ defines the contribution of the cold component to the total spectrum:
\be
       F_{\rm tot} = f_1 \times F_1 + (1-f_1) \times F_2.
\ee
To first approximation, $f_1$ also defines the relative observed NS area occupied by the cold component.
More specifically, the spot sizes depend to a certain extent on the chosen inclination angle $i$ and the co-latitude of the first 
spot center $\theta_{\rm sp}$; see details in \citet{2017A&A...600A..43S}.

\begin{table}
    \centering
    \begin{tabular}{llcc}
       \hline
          \hline
    Parameter  & Unit & $carbatm$ & $hatm$ \\
    \hline
        $N_{\rm H}$ & $10^{22}$\,cm$^{-2}$ &  2.29 & 2.43 \\
        $T_1$ & MK                    &  1.40 & 1.70\\
        $T_2$ &  MK                             & 2.30 & 4.25 \\
        $f_1$ &                                 & 0.51 & 0.977\\ 
        \hline
    \end{tabular}
    \caption{Parameters of the two NS  atmosphere models with nonuniform surfaces. 
    }
    \label{tab1}
\end{table}

\begin{figure}
\centering
\includegraphics[width=0.85\columnwidth]{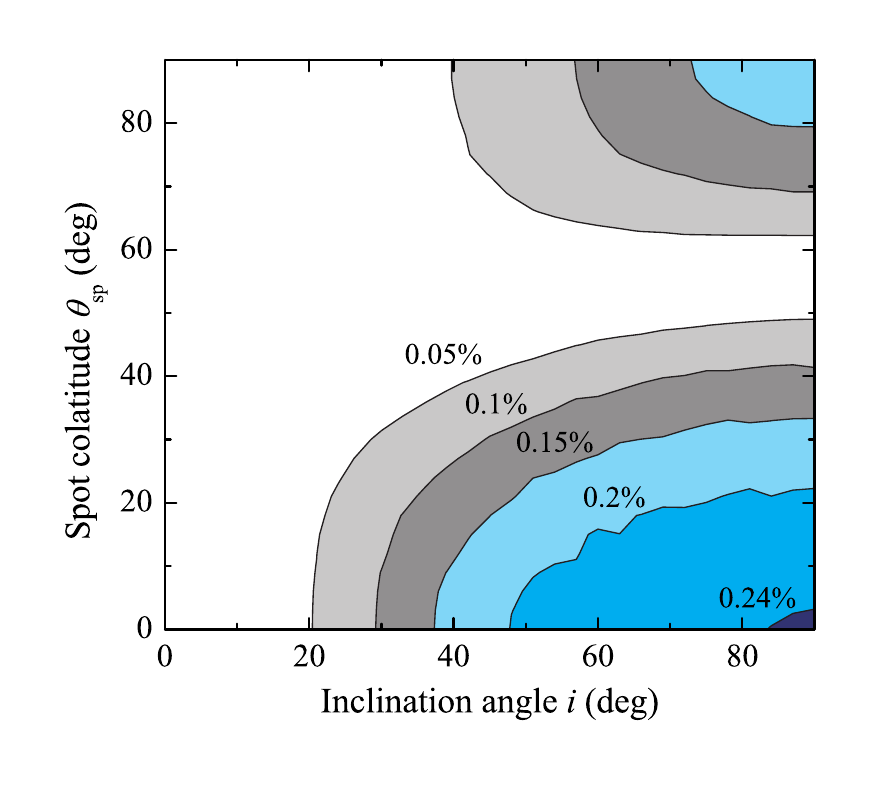}
\caption{Map of the phase-averaged PD in the 2--8\,keV band depending on the inclination angle and the
colatitude of the spot center $\theta_{\rm sp}$ computed for the model with the carbon spot $carbatm$.
The curves of equal PD are labeled.
}
\label{fig:pdmapC}
\end{figure}

\begin{figure}
\centering
\includegraphics[width=0.85\columnwidth]{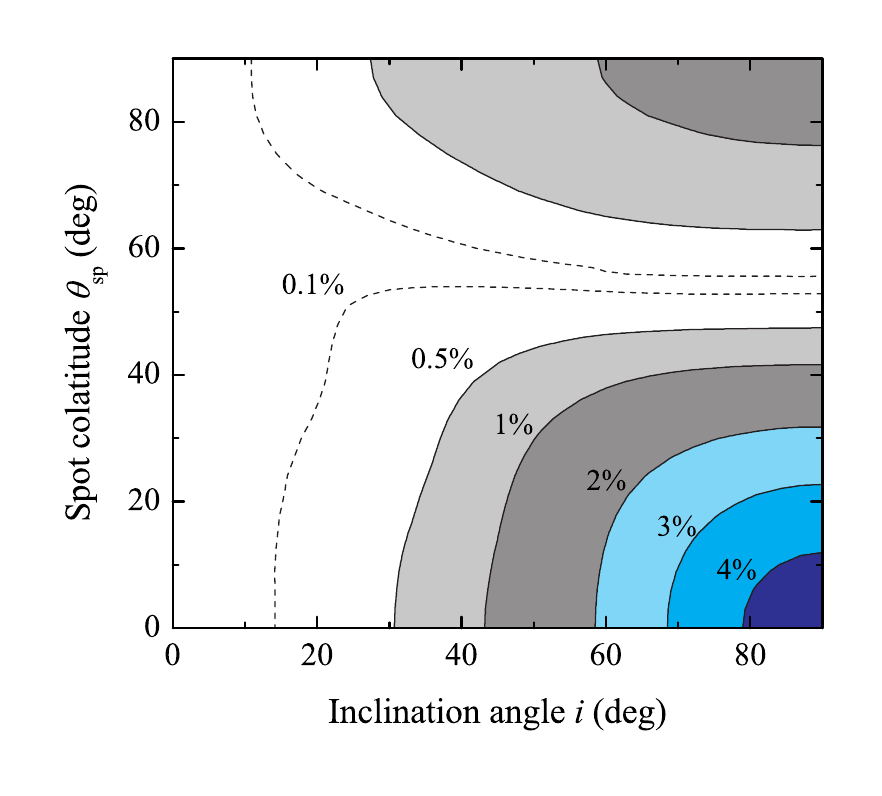}
\caption{Same as Fig.\,\ref{fig:pdmapC} but for the model with the hydrogen spot $hatm$.
}
\label{fig:pdmapH}
\end{figure}

\begin{figure}
\centering
\includegraphics[width=0.85\columnwidth]{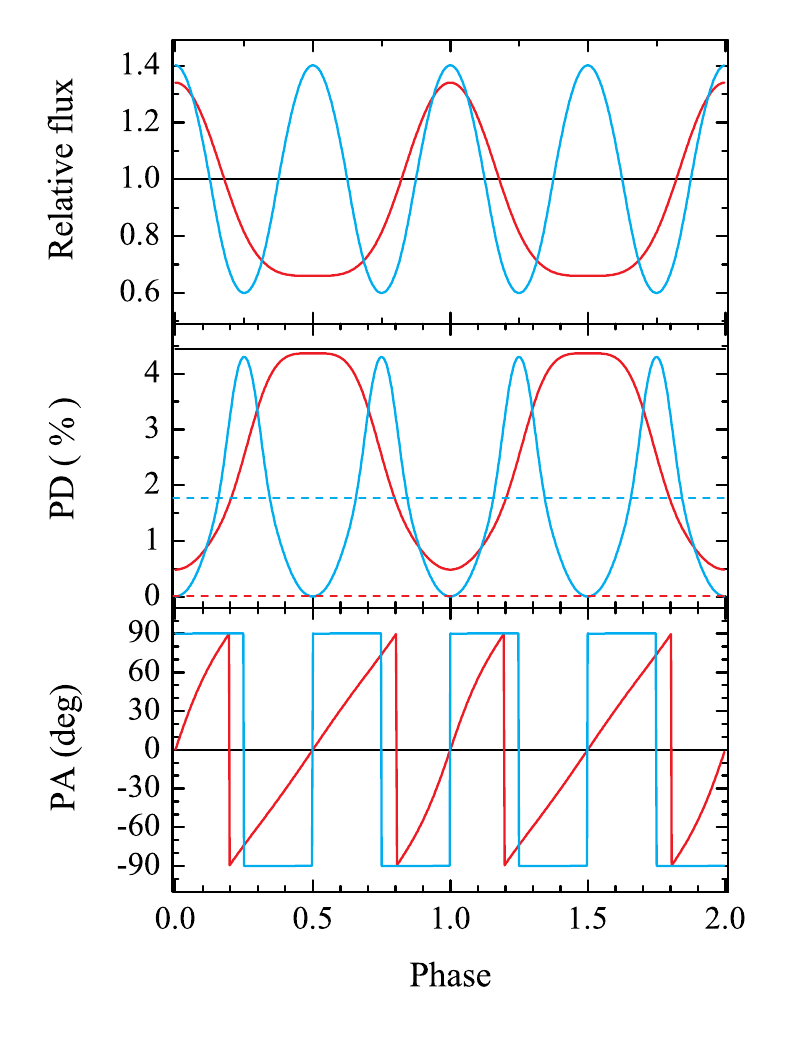}
\caption{Relative pulse profiles (top panel), the phase-dependent PD (middle panel), and the phase-dependent PA (bottom panel) computed in the \textit{IXPE}  energy band 2--8~keV for three sets of angles: the spot center colatitude $\theta_{\rm sp}=0\degr$ and the inclination angle $i=90\degr$ (black lines), $\theta_{\rm sp}=90\degr$, $i=90\degr$ (blue curves), and $\theta_{\rm sp}=60\degr$, $i=30\degr$ (red curves). 
The phase-averaged PDs are shown with dashed lines. The global model $hatm$ (see Table~\ref{tab1}) was used. 
}
\label{fig:phaseH}
\end{figure}

\begin{figure}
\centering
\includegraphics[width=0.85\columnwidth]{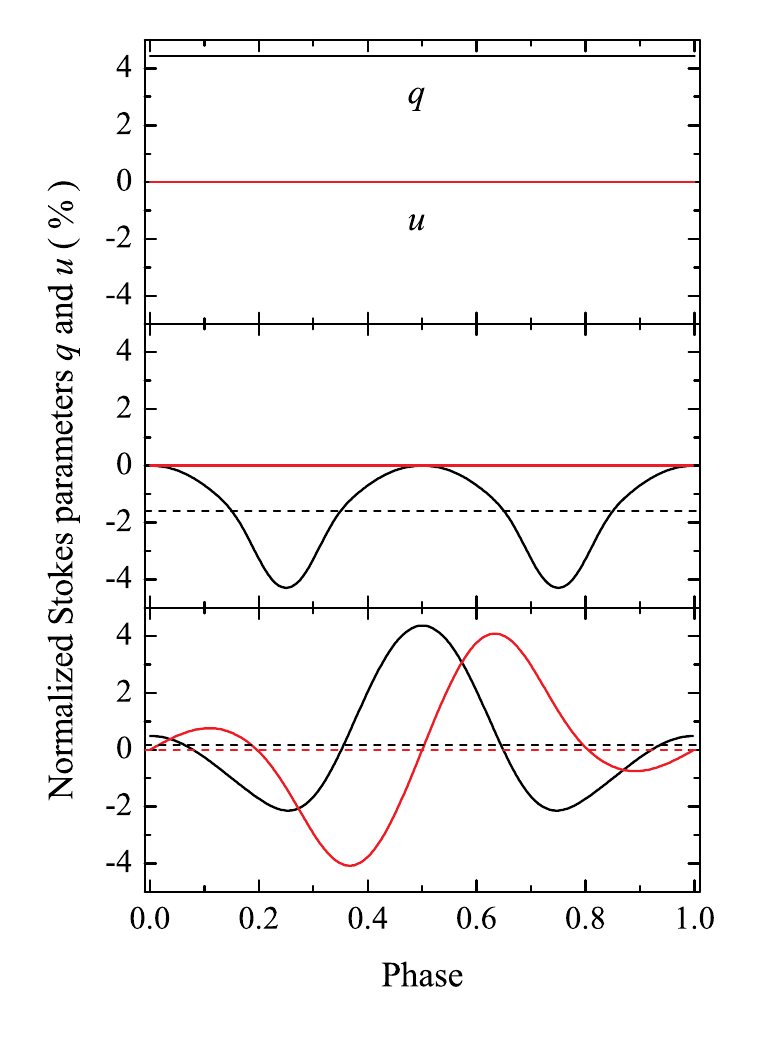}
\caption{Phase-dependent normalized Stokes parameters 
 $q=Q/F$ (black curves) and $u=U/F$ (red curves) 
for the same three models as presented in Fig.\,\ref{fig:phaseH}.
The top panel corresponds to angles $\theta_{\rm sp}=0\degr$, $i=90\degr$, the middle panel to $\theta_{\rm sp}=90\degr$, $i=90\degr$, 
and the bottom panel to angles $\theta_{\rm sp}=60\degr$, $i=30\degr$. 
The phase-averaged values $q$ and $u$ are shown with the dashed lines.
}
\label{fig:phaseQU}
\end{figure}

\begin{figure}
\centering
\includegraphics[width=0.85\columnwidth]{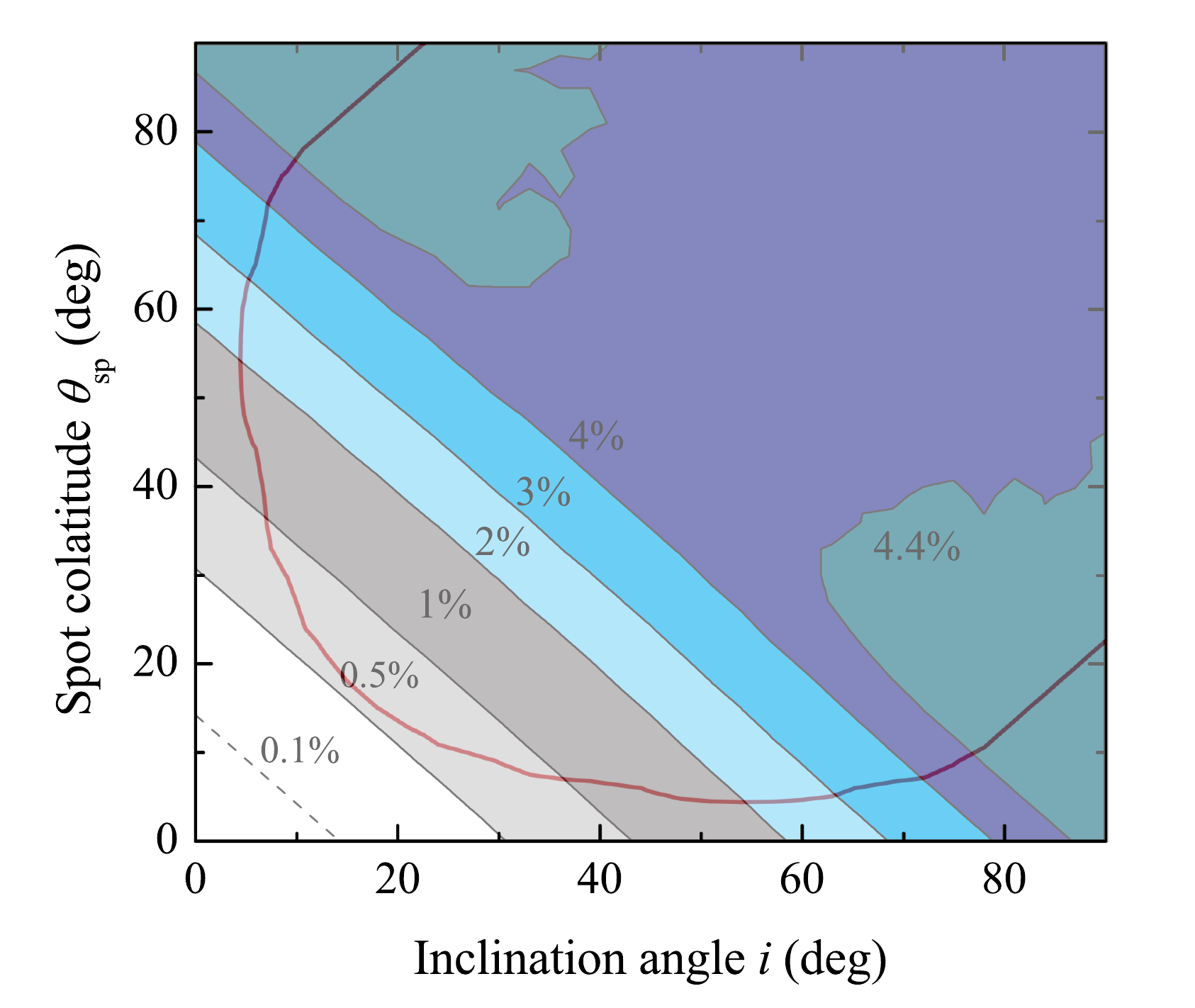}
\caption{Map of the maximum PD in the 2--8~keV band, achievable at some rotational phases, 
depending on the inclination angle and the co-latitude of the spot center $\theta_{\rm sp}$ computed for 
the model with the hydrogen spot $hatm$. The robust observational limit on the pulsed fraction of 9.7\% is also
shown with a thick red curve \citep[see details in ][]{2022NatAs.tmp..224D}. The curves of equal PD are labeled.
}
\label{fig:pdHmax}
\end{figure}

For these parameters, the surface gravity is $\log g \approx 14.2$. There are no models computed for that $\log g$, and we used model atmospheres with the closest value, of namely $\log g = 14.15$. We computed the local PD$_E$ for five 
angles $\alpha$ ($\mu=\cos\alpha$ are the same as presented in Fig.\,\ref{fig:catm2mk}) for two-component   carbon and two-component hydrogen model atmospheres with the effective temperatures presented in Table\,\ref{tab1} (in both cases, components correspond to emission from the hot spot and cooler remaining part of the NS surface). 

The values of PD$_E(\mu)$ together with the data regarding the angular distribution of the emerging radiation computed previously  
\citep{2017A&A...600A..43S} were used for computation of the PD in the \textit{IXPE}  energy band 2--8~keV for models $carbatm$ and $hatm$ for all possible angles 
$i$ and $\theta_{\rm sp}$. 
Maps of the obtained phase-averaged PD are shown in Figs.\,\ref{fig:pdmapC} and \ref{fig:pdmapH}.
The maximum PD in both cases is expected at the angles $i \approx 90\degr$, $\theta_{\rm sp} \approx 0\degr$, when the rotation axis is almost normal to the line of sight, whereas the hot spots are situated near the rotation poles.
The secondary, smaller maximum corresponds to the angle $\theta_{\rm sp} \approx 90\degr$, when the hot spots are located on the NS equator,
and for the same inclination of the observer at $i \approx 90\degr$.  

The expected maximum polarization $\approx$4.4\% for the $hatm$ model is about 20 times larger than for the model $carbatm,$ although the local polarization is higher for the carbon model atmospheres (see previous two subsections).
The obvious reason is the small hot spot sizes in the $hatm$ model ($\approx 13\degr$ vs. $\approx 60\degr$). 
In the case of large spots, the polarization from the opposite parts of the spots compensate each other more effectively. 
 
We also present model predictions for variations of the flux, the PD, and the PA with pulse phase for three pairs of angles (case 1: $\theta_{\rm sp}=0\degr$ and $i=90\degr$; case 2: $\theta_{\rm sp}=90\degr$ and $i=90\degr$; case 3: $\theta_{\rm sp}=60\degr$ and  $i=30\degr$) for the $hatm$ model; see Fig.\,\ref{fig:phaseH}. 
These examples demonstrate that the PD at some pulse phases  can be as high as the maximum phase-averaged PD at the angles $\theta_{\rm sp}=0\degr$ and $i=90\degr$.
The highest PD is reached at the minimum fluxes when the spots are located at the edge of the observed NS hemisphere. 
The corresponding values for the normalized Stokes parameters $q=Q/I$ and $u=U/I$ as a function of the rotation phase are shown  in Fig.\,\ref{fig:phaseQU}; their behavior with the rotation phase explains why the PD averaged over pulse phase is different for cases  2 and 3 despite the similar amplitude of PD variation over the pulse phase.
 
Considering that polarization properties appear to be strongly variable with pulse phase regardless of the model used, detection of the polarization at some particular pulse phase might in principle be easier than in the phase-averaged case. 
We therefore also computed a map of the maximum PD at different rotation phases (Fig.\,\ref{fig:pdHmax}). 
It is clear that the detectability of sufficiently high polarization in separate phase bins could indeed be promising, with a caveat that no pulsations have been detected in any of the CCOs in question. 

Figure\,\ref{fig:pdHmax} also shows a curve corresponding to the computed pulsed fraction $PF=(F_{\rm max}-F_{\rm min})/(F_{\rm max}+F_{\rm min}) = 9.7$\%.
This value is a strict upper limit  on the possible pulsations of the X-ray flux of the CCO in HESS J1731$-$347; see details in \citet{2022NatAs.tmp..224D}. 
The regions of the angles to the left and below the curve correspond to  $PF< 9.7\%$ and these angle combinations ($i, \theta_{\rm sp}$) are allowed if  the assumed model with two hot hydrogen spots is correct.  
We therefore calculate a $\approx 98.5\%$ probability that there is a phase interval where PD exceeds 1\%.

\section{Discussion and conclusions}

We investigated the expected X-ray polarization properties of nonpulsing CCOs assuming that their surfaces are nonuniform and covered by pure carbon or hydrogen envelopes with two symmetric hot bright spots on the surface. 
We also assumed that the magnetic field on the NS surfaces is weak and can be neglected.
 
First of all, we calculated the local polarization properties of pure carbon and pure hydrogen NS atmospheres with low effective temperatures ($T_{\rm eff} \sim $ 1--6\,MK), assuming that polarization arises due to electron scattering only. 
We found that the local polarization is weak at low photon energies ($<$1\,keV) where true opacity (free-free and bound-free transitions) dominates, and very close to a pure electron-scattering atmosphere at high photon energies ($>$10--50\,keV) where electron scattering dominates. 
However, only the first case is of physical relevance as there is no detectable X-ray emission above $\sim$10\,keV from any of the CCOs. 
Between these two regions, there is an energy band where true opacity and electron scattering contribute comparable amounts to the total opacity. 
The PD of the emergent radiation in that band exceeds that of a pure electron-scattering atmosphere and can reach, at the largest zenith angles, up to 15\%--25\% for hydrogen atmospheres and 30\%--40\% for carbon atmospheres. 
The PD is higher for the low-temperature models and weakly depends on the surface gravity. 
This kind of amplification  of the polarization at comparable amounts of true opacity and electron scattering was found by \citet{1979Ap.....15..162L} and our computations confirm their results.

We also studied the polarization properties of the NS models with nonuniform surfaces that describe the observed X-ray spectrum of the CCO in HESS J1731$-$347 \citep{2022NatAs.tmp..224D}, assuming standard values of mass and radius of the NS  ($M = 1.4$\,$M_\odot$ and $R = 12$\,km). 
We found that the maximum  expected PD in the \textit{IXPE}  band 2$-$8\,keV is low ($< 0.25\%$) for the model with the carbon surface and two bright spots. The spot sizes are substantial (about 60\degr) in that model, and taken together with a relatively small difference in the  temperatures between the spots and the rest of the NS surface (the polarization of the cooler models is higher), these lead to the resulting weak polarization. 
This expected polarization level is barely detectable with \textit{IXPE}.
Indeed, the estimated \textit{IXPE}  count rate while assuming the spectral parameters reported in \cite{2022NatAs.tmp..224D} is $\sim0.1$~count\,s$^{-1}$ and detecting polarization at the level of $\sim0.25$\% would require exposure of close to 1 gigasecond, even while neglecting the contribution from the background. 

On the other hand, the model with a hydrogen surface and relatively small spot sizes ($\sim$13\degr) demonstrates a relatively high maximum PD of up to 4.4\% for some geometries. 
This is already potentially detectable with \textit{IXPE}  with a $\sim$1~Ms exposure or future more sensitive facilities like \textit{eXTP}  \citep{2016SPIE.9905E..1QZ}, although it should be borne in mind that such a high PD is only reached when the NS spin axis is almost perpendicular to the line of sight and the hotspot is close to the spin axis.  

Periodic pulsations of the X-ray flux of the CCO in HESS J1731$-$347 were not found with a strict upper limit of 9.7\% \citep{2022NatAs.tmp..224D}, meaning that the considered model is appropriate if the inclination angle $i$ or the spot colatitude $\theta_{\rm sp}$ (or both) are sufficiently close to 0\degr. 
The phase-averaged polarization is highest at  $\theta_{\rm sp} \approx 0\degr$ and $i$ approaching 90\degr. 
The probability of finding a phase-averaged PD of above 1\% is about 25\%. This corresponds to a required \textit{IXPE} exposure of 100~Ms, which is not realistic. 
In the opposite case, with $\theta_{\rm sp} \approx 90\degr$ and $i$ close to 0\degr, the PD has the same high values at all rotational phases. 
If a periodic PD increase can also be found, the probability of finding a polarization signal at least in some phases due to favorable geometry increases to about 98\%; however, the observation time will also need to be increased accordingly to ensure that sufficient exposure is also accumulated  within individual phase bins (provided that the phase can be determined to begin with). 

The main conclusion of this work is that  PDs of up to a few percent in the CCO in HESS J1731$-$347 can be expected for a hydrogen atmosphere and favorable orientation of the NS. 
Such polarization could be detectable with \textit{IXPE} and future X-ray polarimetry missions within a reasonable exposure time of a few million seconds. 
For a carbon atmosphere, the expected polarization is lower, and is not detectable with current facilities and is unlikely to be detectable with the planned future facilities.
Thus, a detection of polarization from the considered object or other nonpulsating CCOs would be a strong argument in favor of the NS model covered by a hydrogen envelope with two bright spots. 
On the other hand, if the polarization from nonpulsed CCOs is not detected, this would support the hypothesis that the observed emission comes from an almost uniformly emitting NS covered by a carbon envelope, although this nondetection would clearly not be definitive evidence.

Here, we did not consider the possibility that the magnetic field is strong at the bright spots, that is, about $10^{12}$\,G, as was suggested for the CCO in Kes\,79 \citep{2014ApJ...790...94B}. 
The corresponding PD could be much larger in this case. 
The corresponding computations are nevertheless beyond the scope of the current work and will be presented in a separate paper.

\begin{acknowledgements}  
This research has been supported by Deutsche Forschungsgemeinschaft (DFG, grant WE 1312/53-1) and the Academy of Finland grant 333112. 
We thank the German Academic Exchange Service (DAAD, project 57525212) and the Academy of Finland (project 349144) for travel grants.
\end{acknowledgements}

\bibliographystyle{aa}
\bibliography{pol_cco}

\end{document}